%% file: efimov.tex
\documentclass[aps,prd,twocolumn,tightenlines,preprintnumbers,showpacs,superscriptaddress,notitlepage,nofootinbib,eqsecnum,floatfix,longbibliography,10pt]{revtex4-1}

\input{preamble}

\input{def}

\input{affiliations}

\newcount\hour \newcount\hourminute \newcount\minute
\hour=\time \divide \hour by 60
\hourminute=\hour \multiply \hourminute by 60
\minute=\time \advance \minute by -\hourminute

\begin{document}

\title{Determining universal spectra from probability distributions}

\author{Charles Kacir}
\affiliation{\unc}

\author{Joseph Moscoso}
\affiliation{\umd}
\affiliation{\unc}
\affiliation{\lblnsd}

\author{Amy~Nicholson$^*$}
\affiliation{\unc}
\affiliation{\lblnsd}

\author{Thomas R. Richardson}
\affiliation{\ucb}
\affiliation{\lblnsd}

\author{Cade Rodgers}
\affiliation{\yale}
\affiliation{\unc}

\date{\today}

\begin{abstract}
The probability distribution of a two-particle correlation function computed over background auxiliary field configurations, used to generate the interactions, has been shown to inform about the spectra of universal $n$-body clusters~\cite{Nicholson:2012zp}. Here, we utilize two approaches, a numerical lattice computation and an analytic expansion in the limit of large numbers of identical species, in an attempt to refine the initial predictions. Exploratory calculations in these directions are presented, and future investigations laid out. \\

$^*$\textit{speaker}
\end{abstract}

% make title
\maketitle
\preprint{}

%-------------------------------------------------------------------------------
%  Introduction
\section{Introduction\label{sec:intro}}
Lattice methods for computing properties of strongly-interacting systems have been instrumental in the quantitative understanding of theories ranging from Quantum Chromodynamics (QCD) to low-energy Effective Field Theories (EFTs) of atomic and condensed matter systems. However, formulations for theories involving certain physical scenarios, such as the introduction of non-zero baryon chemical potential within lattice QCD, are known to exhibit sign problems, in which the action to be sampled using Monte Carlo methods become complex. The physical mechanisms underlying these sign problems typically appear as exponentially poor signal-to-noise problems in associated correlation functions sampled using a sign-problem free action~\cite{Lepage:1989hd,Parisi:1983ae,Endres:2011mm}.

It has been shown in certain situations that a signal-to-noise problem can be reformulated into a theory having instead a statistical overlap problem~\cite{Endres:2011jm,Endres:2011mm,Grabowska:2012ik}\footnote{Using the standard reweighting method for circumventing a sign problem tends to result in both a symmetric distribution with exponentially small mean in computing the expectation value of the phase, as well as a statistical overlap problem when computing the observable.}. While sign or signal-to-noise problems are associated with nearly symmetric distributions having an exponentially small mean compared to its variance, a statistical overlap problem instead has a largely positive, but highly asymmetric distribution. In practice, this can require an exponentially large number of statistical samples in order to satisfy the Central Limit Theorem (CLT) and faithfully reproduce the mean. 

However, it has been noted in the literature that log-normal distributions seem to be ubiquitous in lattice calculations involving a wide variety of physical scenarios~\cite{DeGrand:2012ik,Endres:2011jm,Grabowska:2012ik,Abbott:2023coj}. While observables having long-tailed distributions are in general difficult to sample, if the distribution is nearly log-normal it becomes possible to sample instead the logarithm of the observable, which obeys a nearly Gaussian distribution. This provides a potential path forward toward resolving noisy systems.

The ``universality" of the log-normal distribution can be understood to stem from a similar argument to the CLT. For the CLT, the sum of values (average) drawn from a random distribution ``flows" toward a normal distribution as the number of draws becomes large. If, rather than summing the randomly drawn values, one takes the product of the logarithm of each value, one obtains the log-normal distribution in the same limit. When computing lattice observables, propagators are formed roughly by taking products of random variables on each time slice. Thus, in the large time limit, one might expect log-normal distributions to appear. This argument is not mathematically rigorous, however, as the propagator is a more complicated set of matrix products stemming from taking the inverse of the fermion matrix. 

Although it has been common knowledge for some time that there are physical mechanisms underlying the moments of statistical distributions of correlation functions~\cite{Parisi:1983ae,Lepage:1989hd,Beane_2011,Beane_2015,Endres:2011mm,Wagman:2016bam}, it has been more difficult to discern whether this problem can be inverted to extract physical information by studying the distributions directly.
Does the universality of the log-normal distribution have anything to do with universal physics? This remains an open question. However, it is known that in certain cases, the moments of the probability distribution of a fermion correlator can be exploited to understand the spectrum of the theory~\cite{Nicholson:2012zp,Abbott:2023coj}. 

The argument is as follows: for a lattice computation involving matter fields interacting via exchange of bosonic fields, a correlation function for some operator is computed by first integrating out the matter degrees of freedom by hand. This results in a function of the matter propagators, computed on a given background configuration for the bosonic fields. For example, given matter fields, $\psi(x)$, and bosonic interaction fields, $\phi(x)$, a correlation function, $C_1(t)$, for a single particle propagating from Euclidean time $\tau=0$ to $\tau$ is given by,
\begin{eqnarray}
C_1(\tau) = \sum_{a=1}^{N_\mathrm{cfg}} M[\phi_a,\tau]   \ , 
\end{eqnarray}
where the set of $N_\mathrm{cfg}$ background fields, $\{\phi_1, \phi_2, \cdots \phi_{N_\mathrm{cfg}}\}$, is generated using the probability measure, $\rho[\phi] = \left(\det \mathcal{K} \right)^{\pm N} e^{-S_{\phi}}$, where $N$ denotes the number of degenerate species of matter particles and the $\pm$ depends upon the choice of either fermionic or bosonic matter fields. It is assumed that the original action of the field theory takes the form, $S = S_{\phi} + \bar{\psi}\mathcal{K}\psi$, with bosonic action $S_{\phi}$. Then we can define the propagator, $M[\phi_a]$ to be the operator $\mathcal{K}^{-1}[\phi_a]$, projected onto some choice of initial and final states. For $N\geq n$, an $n$-body correlation function involving only non-identical species can take the form,
\begin{eqnarray}
\label{eq:moments}
C_n(\tau) = \sum_{a=1}^{N_{\mathrm{cfg}}} \left(M[\phi_a,\tau]\right)^n   \ .
\end{eqnarray}

In the limit of large Euclidean time, a spectral decomposition of the correlation functions gives,
\begin{eqnarray}
\label{eq:specdec}
    C_n(\tau)  \underset{t\to \infty}{\longrightarrow} A^{(n)}_0 e^{-E^{(n)}_0 \tau} \ ,
\end{eqnarray}
where $A^{(n)}_0,E^{(n)}_0$ are the overlap and ground state energy, respectively, of the lowest state containing $n$ matter fields. Noting that the right-hand side of Eq.~\ref{eq:moments} is equivalent to the $n$th raw sample moment of the observable $M[\phi,\tau]$ with density, $\rho[\phi]$, we finally have our connection between moments of a distribution and spectra. A caveat, however, is related to the number of degenerate species in the theory: only if there are at least $N\geq n$ species in the ``vacuum" (the partition function, $\rho[\phi]$), or if the spectrum is independent of this vacuum number, can we determine energies for a physical $n$-body state. 

While the issue of computing individual moments, whether numerically or analytically, is equivalent to solving a given $n$-body system exactly, an advantage can be gained if the probability distribution is perturbatively close to some known analytic form. Given the pervasive appearance of the log-normal distribution in lattice computations, this seem a natural starting point. 

A promising class of theories which may satisfy the above criteria are those involving Efimov clusters. In Ref.~\cite{Nicholson:2012zp}, it was shown that the $n$th moment of the probability distribution of a two-particle correlation function, in a theory where the two-body interaction, generated via exchange of auxiliary bosonic fields, is tuned such that the scattering length of the system approaches infinity, is related to the binding energy, $-E_{2n}$, of a universal $2n$-body cluster tied to an Efimov trimer. Numerical evidence that this distribution is approximately log-normal was presented, giving a relation, $E_{2n}/E_4 = 1/2n(n-1)$, in the limit that the distribution is exactly log-normal. 

\begin{figure}
\includegraphics[width=\linewidth]{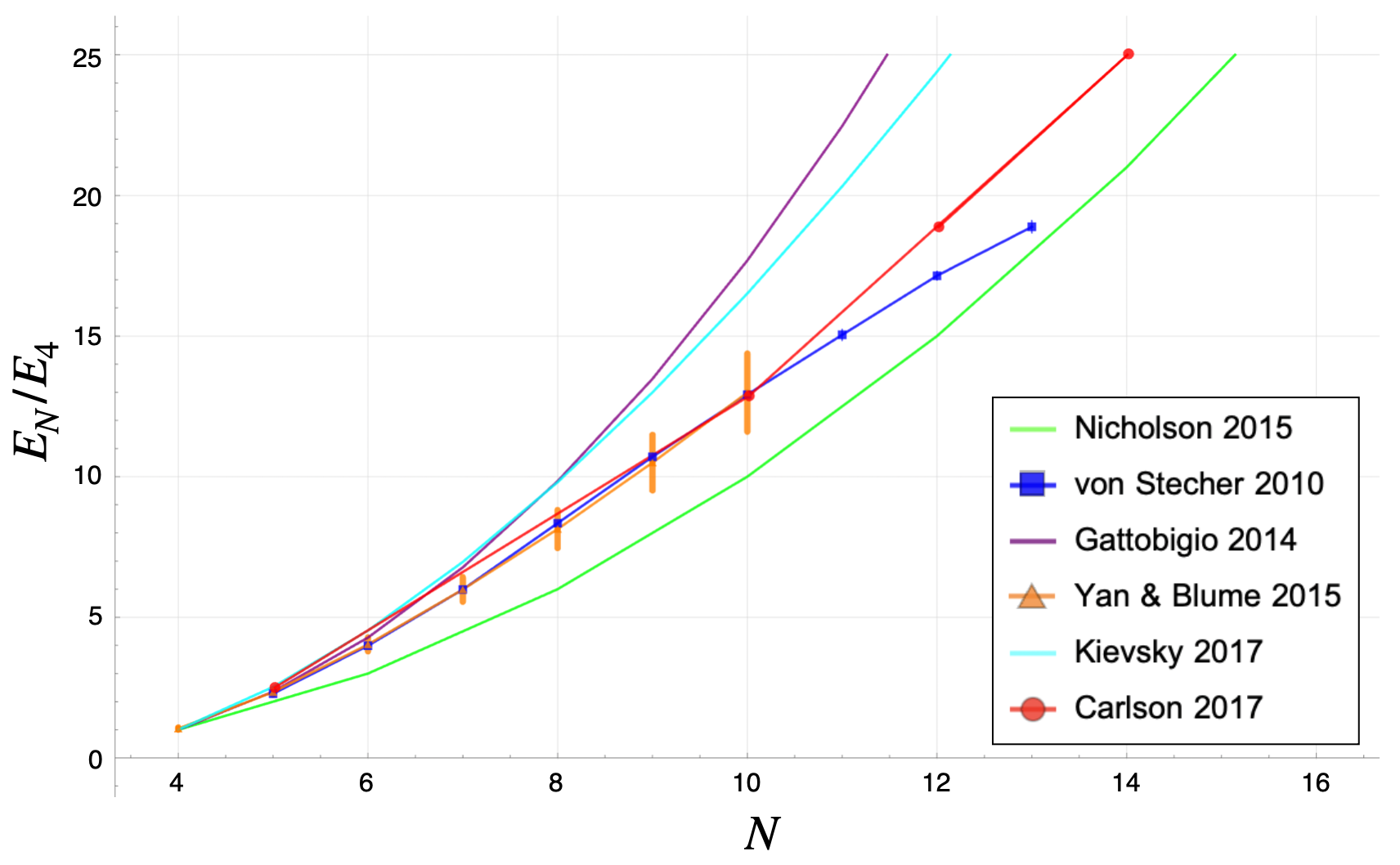}
\caption{Energy of a universal $N$-body cluster, $E_N$, relative to the 4-body energy, $E_4$, as a function of $N$. Results assuming a log-normal distribution for the $N=2$ correlator are shown in green. Data and other analytic predictions are taken from: von Stecher 2010 ~\cite{vonStecher:2009qw}, Gattobigio 2014 ~\cite{Gattobigio:2013yda}, Yan \& Blume 2015 ~\cite{Yan:2015tla}, Kievsky 2017 ~\cite{Kievsky:2017mjq}, Carlson 2017 ~\cite{Carlson:2017txq}.
\label{fig:Ensummary}}
\end{figure}

In Figure~\ref{fig:Ensummary}, we show a summary of direct calculations of the $n$-body energies in the literature, as well as the log-normal prediction. As best estimates from direct calculations are systematically higher than the log-normal prediction, physical deviations from a purely log-normal distribution may be present. Here, we attempt to quantify any such corrections. 

In what follows, we explore two approaches which utilize approximately log-normal distributions to determine spectra. First, we apply a numerical lattice formulation to directly measure sample moments of the logarithm of a correlation function. In this lattice formulation, the spectrum is independent of the number of species in the vacuum, and therefore satisfies our criteria for physical systems. In the second method, we use analytic field theory to study the probability distribution in the limit $N\to \infty$, where again our criteria $N \geq n$ is satisfied.
Both of these techniques use as their starting point a perturbative expansion around a log-normal form.

\section{Lattice formulation}

We study a system of particles interacting via exchange of bosonic auxiliary fields,
\begin{eqnarray}
\mathcal{L} = \psi^{\dagger}\left(\partial_{\tau}-\frac{\nabla^2}{2M} +\phi\right)\psi + \frac{\phi^2}{2g} \ ,
\end{eqnarray}
where the coupling, $g$, is chosen such that the two-body interactions between any pair of particles is tuned to unitarity. We use the discretized formulation of Ref.~\cite{Endres:2012cw}. Note that in this formulation the determinant of the operator sandwiched between matter fields is constant, and does not contribute to importance sampling. Therefore, the particle content of the theory, namely, whether the particles are bosons or fermions, and the number of identical species, is encoded only in the observable, and does not affect the partition function, $\rho[\phi] \propto e^{-\frac{\phi^2}{2g}}$.  

The observable we compute is $C_2(\tau)$; this choice ensures that the operator on the right-hand side of Eq.~\ref{eq:moments} is positive definite, a requirement for log-normal distributions, and to help minimize the noise entering into the computation of moments from finite sample sizes. Inspection of Eq.~\ref{eq:specdec} implies that raw moments of this operator correspond to the energies of $2n$-body clusters of non-identical particles. The coupling is chosen to reproduce the unitary limit, and it is known that in this limit $2n$-body clusters, at least for relatively small $n$, are tied to Efimov trimers obeying a log-periodic scaling symmetry~\cite{vonStecher:2009qw,Gattobigio:2013yda,Yan:2015tla,Kievsky:2017mjq,Carlson:2017txq}.

We note that there is no explicit three-body term in this Lagrangian; this does not imply that there is no three-body interaction, it simply limits our freedom to choose this interaction by tuning a coupling. The three-body interaction is set, instead, by the particular discretization chosen, as well as the cutoff (lattice spacing). Given the log-periodic dependence of the three-body coupling on the cutoff, whether a particular discretization results in an attractive or a repulsive effective three-body force is not \textit{a priori} known. One must check whether the cutoff scale is sufficiently far below the ground-state energy to give universal behavior. However, this last statement is true for any lattice (really, any type of numerical) computation: one must study and either remove non-universal cutoff effects or show that they are negligible within the quoted error bars. Because the ratio of energies, $E_{2n}/E_4$, should be independent of the cutoff, we can systematically remove non-universal effects from this quantity to extract universal physics. While this may be simpler to accomplish in certain cases through tuning of an explicit three-body coupling, it is not necessary.

Deviations from a purely log-normal distribution are most easily quantified through a cumulant expansion of the logarithm of the observable, which gives the relation:
\begin{eqnarray}
\label{eq:cmexp}
    \ln C_2^{(N_\kappa)}(\tau) = \sum_{n=1}^{N_\kappa}\frac{\kappa_n(\tau)}{n!} \underset{\tau,N_\kappa \to \infty}{\longrightarrow} -E_2 \tau\ ,
\end{eqnarray}
where $\kappa_n(\tau)$ is the $n$th cumulant of $\ln\left( M[\phi,\tau]\right)^2$, and a spectral decomposition has been used to take the final limit. Given the tuning to unitarity, we expect $E_2 = 0$ once discretization effects are removed. If the distribution is log-normal, we furthermore expect $\kappa_{n>2}(\tau) = 0$, and the expansion is cut off exactly at second order. Deviations from log-normal, therefore, result in non-zero contributions to $\kappa_{n>2}(\tau)$. However, one must first extrapolate the results to infinite lattices to remove any discretization effects and isolate purely physical deviations from log-normal. 

\begin{figure}
\includegraphics[width=\linewidth]{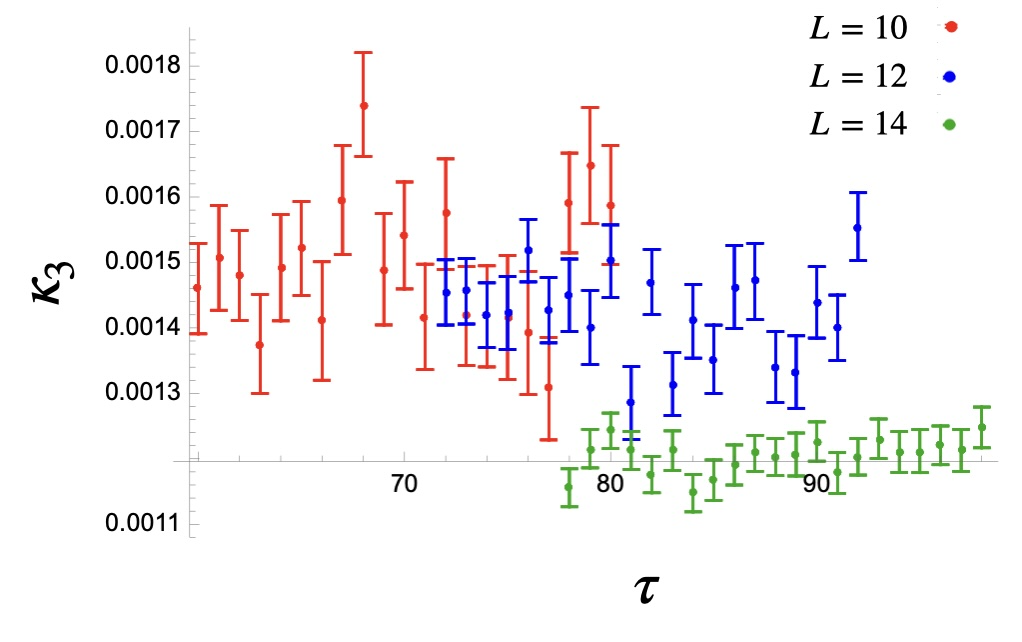}
\includegraphics[width=\linewidth]{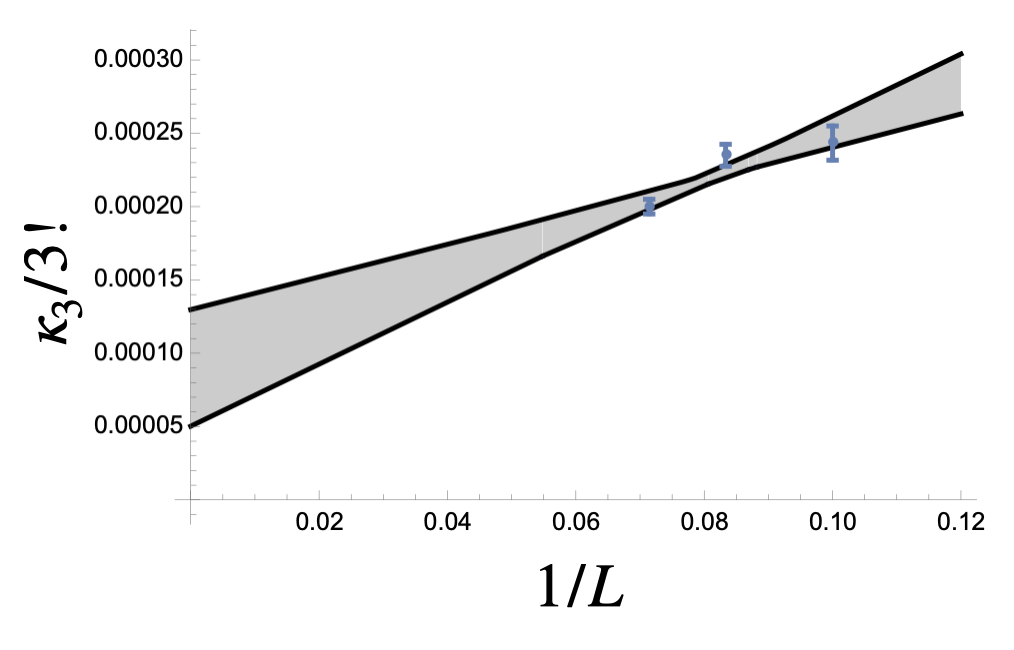}
\caption{Contributions to the third cumulant, $\kappa_3$, of $\ln \left(M[\phi,\tau]\right)^2$, as a function of $\tau$ (top) and $1/L$ (bottom). The band on the lower figure corresponds to a linear extrapolation in $1/L$. 
\label{fig:kappa3}}
\end{figure}

Such a study is underway, utilizing lattices having different numbers of spatial sites, $L^3$, as well as different tunings of the action~\cite{Endres:2012cw}. Results for $\kappa_3$ as a function of $\tau$ and $1/L$ are shown in Fig.~\ref{fig:kappa3}. For the time windows considered, $\kappa_3(\tau)$ appears to have plateaued, indicating a lack of contamination from excited states. Preliminary linear extrapolations in $1/L$ seems to indicate that the contribution to the cumulant expansion from $\kappa_3$ is extremely tiny, of $\mathcal{O}(0.01\%)$. More thorough investigations are underway.

Once the extrapolated $\kappa_{n>2}$ have been quantified, one must then determine how these cumulants of $\ln \left(M[\phi,\tau]\right)^2$ affect the distribution of $\left(M[\phi,\tau]\right)^2$, in order to determine the sizes of their contributions to the spectrum. Machine learning techniques for this matching are currently under investigation. 

\section{Analytic QFT technique}

In the continuum, the 2-fermion correlation function discussed above corresponds to the path integral
\begin{eqnarray}
\label{eq:C2phi}
C_2(\tau) = \langle C_2(\phi,\tau) \rangle \equiv \int \mathcal{D} \phi e^{-S[\phi]} C_2(\phi,\tau) \ ,
\end{eqnarray} 
where $S[\phi] \equiv \int d^3x \int d\tau \left(\frac{\phi^2}{2g} \pm \mathrm{Tr} \ln \mathcal{K}[\phi]\right)$, with $\mathcal{K}$ the operator corresponding to the matter term in the Lagrangian, $\mathcal{L} = \Psi^{\dagger} \mathcal{K}[\phi] \Psi + \frac{\phi^2}{2g}$. The choice of fermionic or bosonic statistics for the matter fields determines the sign of the second term in $S[\phi]$. 
One can define a probability distribution for an observable, $X$, as,
\begin{eqnarray}
    P[x] =\mathcal{N} \int \mathcal{D}\phi e^{-S[\phi]}\delta\left(X[\phi]-x\right) \ ,
\end{eqnarray}
where $\mathcal{N}$ is a normalization. For this work, we will take $X[\phi] = \ln C_2(\phi,\tau)$, as we expect this distribution to be perturbatively close to a Gaussian distribution. 

As in Ref.~\cite{Grabowska:2012ik}, we avoid the singular delta function by turning to the characteristic function,
\begin{eqnarray}
\label{eq:charfun}
\Phi_{\ln C_2}(s) = e^{-W(s)} = \mathcal{N}\int \mathcal{D}\phi e^{-S[\phi] + i s \ln C_2[\phi,\tau]} \ ,
\end{eqnarray}
where $W(s) = -\sum_n \frac{(is)^n}{n!}\kappa_n$, as given by the cumulant expansion, Eq.~\ref{eq:cmexp}, can be computed using the sum of connected graphs having $2n$ external lines.

For the continuum theory, we cannot in general neglect the term stemming from the determinant of $\mathcal{K}$; therefore, we take the number of identical species of matter fields, $N$, to be large such that $N>n$ is satisfied for an arbitrarily large $n$th moment. One must carefully check whether the Efimov spectrum is affected by $N$. This is an on-going study.  

\subsection{Effective Field Theory at Large N}

We now specialize to fermionic matter fields to simplify the number of ways in which particles can interact via contact interactions. An effective field theory for N-species of interacting, non-relativistic, two-component fermions may be written,
\begin{eqnarray}
\mathcal{L} = \Psi_a^{\dagger}\left(\partial_{\tau}-\frac{\nabla^2}{2M} \right)\Psi_a + g \left(\Psi_a^{\dagger}\Psi_a\right)^2 \ ,
\end{eqnarray}
where $\Psi_a = \left(\begin{array}{c}
\psi_{\uparrow,a} \\
 \psi_{\downarrow,a}
 \end{array}\right)$. The action can be made bilinear in the fermion fields by introducing an auxiliary field via Hubbard-Stratonovich transformation. There are two standard formulations, with the auxiliary field acting in either the ``density" channel, with Lagrangian
 \begin{eqnarray}
\mathcal{L}_{\mathrm{den}} = \Psi_a^{\dagger}\left(\partial_{\tau}-\frac{\nabla^2}{2M} +\phi\right)\Psi_a + \frac{\phi^2}{2g} \ ,
\end{eqnarray}
or the ``BCS" channel,
\begin{eqnarray}
\mathcal{L}_{\mathrm{BCS}} &=& \Psi_a^{\dagger}\left(\partial_{\tau}-\frac{\nabla^2}{2M} \right)\Psi_a + \frac{|\phi|^2}{2g}  \cr
&+&  \phi^{*} \psi_{\downarrow,a}\psi_{\uparrow,a}+ \phi \psi^{\dagger}_{\uparrow,a}\psi^{\dagger}_{\downarrow,a}\ .
\end{eqnarray}
For the latter, we use the Nambu-Gorkov basis to bring the action into bilinear form for the fermion fields,
\begin{eqnarray}
\mathcal{L} &=& \tilde{\Psi}^{\dagger}_a \mathcal{K}_{\mathrm{BCS}} \tilde{\Psi}_a + \frac{|\phi|^2}{2g}\ , \cr
\tilde{\Psi}_a &\equiv& \left(\begin{array}{c}
\psi_{\uparrow,a} \\
\psi^{\dagger}_{\downarrow,a}
\end{array}\right)  \ , \cr
\mathcal{K}_{\mathrm{BCS}} &\equiv& \left( \begin{array}{cc} \partial_{\tau} -\nabla^2/2M & \phi \\
\phi^{*} & \partial_{\tau} +\nabla^2/2M 
\end{array} \right) \ .
\end{eqnarray}

For use in standard few-species Monte Carlo calculations, introducing the field $\phi$ in the ``BCS" channel na\"ively introduces a sign-problem and cannot be log-normal due to the complexity of the $\phi$ field itself. However, in the large-$N$ limit, the mean-field solution for the auxiliary field is found to be real and positive~\cite{Romatschke:2023ztk}. We will explore both formulations here. 

For a given process, the two different formulations give rise to different types of graphs, resulting in differences in the $N$-counting for that process. This is due to the different action of the vertices in the two cases: for the density formulation, the auxiliary field annihilates a particle carrying species $a$, then creates a particle having the same species, while in the BCS formulation, the auxiliary field annihilates two particles (with opposite spin) carrying species $a$, then creates a new pair of particles carrying any species. The latter allows for closed fermion loops, with an associated factor of $N$, while the former does not. 

In this work, we will be considering the following two-particle correlation function in a mixed time-momentum basis,
\begin{widetext}
\begin{eqnarray}
\label{eq:C2denbcs}
C_2^{\mathrm{den(BCS)}}(T) = \frac{M}{\mathcal{N} V}\int [d\phi] e^{-S_{\phi}} \langle \mathbf{p} = \mathbf{0}, \tau = 0|\left(\det \mathcal{K}_{\mathrm{den(BCS)}}\right)^{-1} | \mathbf{p} = \mathbf{0}, \tau = T \rangle \ ,
\end{eqnarray}
\end{widetext}
where, 
\begin{eqnarray}
\mathcal{N} &\equiv& \int \mathcal{D}\phi e^{-S_{\phi}} \left(\det \mathcal{K}_{\mathrm{den(BCS)}}[\phi]\right)^{N} \cr
&=& \int \mathcal{D}\phi e^{-S_{\phi} \pm N \mathrm{Tr} \ln \mathcal{K}_{\mathrm{den(BCS)}}} \cr
&\equiv& \int \mathcal{D}\phi e^{-S_{\mathrm{den(BCS)}}}\ . 
\end{eqnarray}
Here,  
$S_{\phi} \equiv \int d\tau d^{3} x \frac{|\phi|^2}{2g}$, and $\mathcal{K}_{\mathrm{den}} \equiv \left(\partial_{\tau}-\frac{\nabla^2}{2M} +\phi\right)\times\mathbbm{1}_{2\times 2}$.

\begin{figure}
\includegraphics[width=\linewidth]{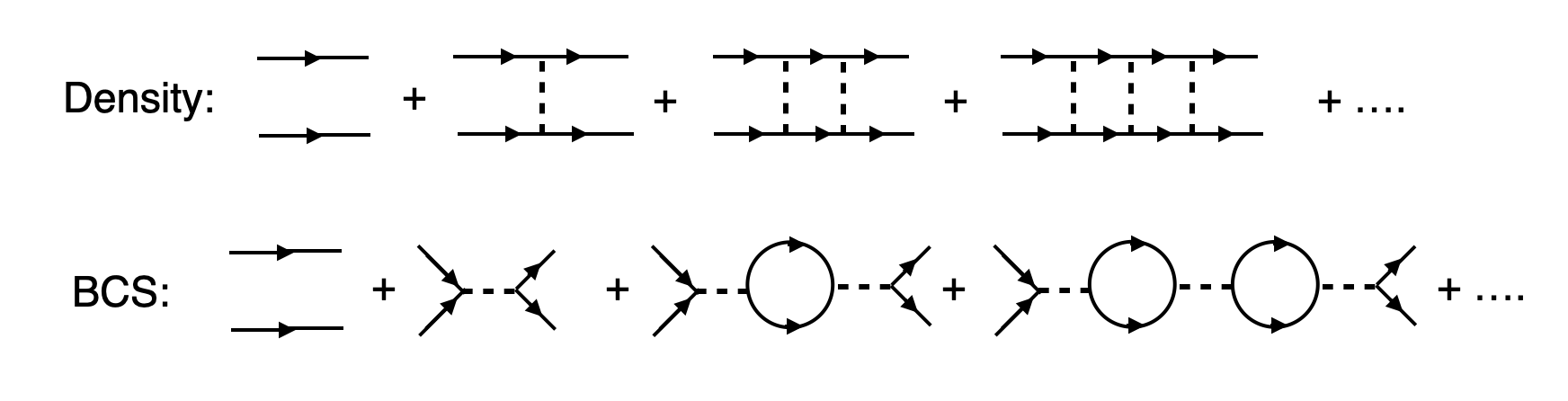}
\caption{ 
\label{fig:2part}Graphs contributing to the two-particle correlation function, Eq.~\ref{eq:C2denbcs}. Fermionic propagators are denoted by solid lines with arrows following the progression of time, while bosonic auxiliary field contributions are dashed. }
\end{figure}

Diagrammatic expansions for the two formulations are shown in Fig.~\ref{fig:2part}. 
In both cases, one may sum the full geometric series of graphs to obtain,
\begin{eqnarray}
\label{eq:C2den}
\hspace{-5mm} C_2^\mathrm{den}(\tau) &=&  M\int \frac{d\omega}{2\pi}e^{-i\omega \tau} \left(1 + g\sum_{n=0}^{\infty} \left(g I_0(\omega)\right)^n\right)\cr 
 &=&  M \int \frac{d\omega}{2\pi}e^{-i\omega \tau} \left(1 + \frac{g}{1-g I_0(\omega)} \right) \ ,
\end{eqnarray}
and
\begin{eqnarray}
\label{eq:C2BCS}
\hspace{-5mm} C_2^\mathrm{BCS}(\tau) &=&  M\int \frac{d\omega}{2\pi}e^{-i\omega \tau} \left(1 + g\sum_{n=0}^{\infty} \left(g N I_0(\omega)\right)^n\right)\cr
&=& M\int \frac{d\omega}{2\pi}e^{-i\omega \tau} \left( 1 + \frac{g}{1-g N I_0(\omega)} \right)  , 
\end{eqnarray}
where,
\begin{widetext}
\begin{eqnarray}
I_0(\omega) &\equiv& \int \frac{d^4q}{(2\pi)^4} \frac{1}{\left(iq_0+i\omega/2-q^2/2M\right)\left(iq_0+i\omega/2+q^2/2M\right)}
= -\frac{M}{4\pi}\left(\mu + \sqrt{i M \omega}\right)\ ,
\end{eqnarray}
\end{widetext}
in the PDS subtraction scheme~\cite{Kaplan:1998tg}.

Note the factor of $N$ which is present for the BCS case but not the density case due to the lack of closed fermion loops in the latter. In fact, there cannot be closed fermion loops in any correlation function within the density formulation, as the number of species participating in any process is set only by the number of distinct species in the external lines for that process. Therefore, no factors of $N$ will appear for calculations within this formulation, and the ``extra" species present in the theory that are not present in external lines are fully decoupled. On one hand, this means that the physics of interest, namely Efimov states, is completely unaffected by the presence of the extra species in the theory. On the other hand, this means that we cannot exploit the large $N$ limit.

The BCS formulation, however, does allow for closed fermion loops and therefore follows a different $N$-counting. Any potential effect on the expected universal behavior of the spectrum is currently under investigation. For now, we simply proceed with studying this distribution in the large $N$ limit to see what we find.

We can absorb a factor of $N$ into the coupling, $\tilde{g} \equiv g N$, such that $\tilde{g}$ is $N$-independent. Then the two-particle correlation function becomes:
\begin{eqnarray}
\label{eq:C2}
C_2(\tau) = M\int \frac{d\omega}{2\pi}e^{-i\omega \tau}\left(1 + \frac{\tilde{g}/N}{1+\frac{\tilde{g} M}{4\pi}\left(\mu +  \sqrt{i M \omega}\right)} \right)\ , \cr
\end{eqnarray}
where we have dropped the BCS label and will only use the BCS formulation from here on.  Ignoring the branch cut and approximating the integrand in Eq.~\ref{eq:C2} near the bound-state pole gives,
\begin{eqnarray}
\label{eq:C2}
C_2(\tau) = \frac{8\pi}{\tilde{g}N M^2}(\tilde{g}M \mu+4\pi) e^{-1/M(\mu+4\pi/(\tilde{g}M))^2\tau} \ .
\end{eqnarray}
Comparing with the expected form for the two-fermion correlation function near unitarity,
\begin{eqnarray}
C_2(\tau) = \mathcal{Z}_0 e^{-1/(a^2 M)\tau} \ ,
\end{eqnarray}
where $a$ is the scattering length, we find
\begin{eqnarray}
\tilde{g} = \frac{4\pi}{M(1/a-\mu)} \ .
\end{eqnarray}
Defining a dimensionless coupling $\hat{g} \equiv \frac{M \mu}{4\pi} \tilde{g}$, we find the beta function 
\begin{eqnarray}
\beta(\mu) &=& \mu \frac{\partial}{\partial \mu} \hat{g} = \mu/(1/a-\mu) [1- 1/(1/a-\mu)] \cr
&=& -\hat{g}(1-\hat{g}) \ ,
\end{eqnarray}
with fixed points $\hat{g} = 0$ corresponding to a non-interacting system, and $\hat{g} = 1$ corresponding to unitarity. 

We now identify our action as,
\begin{eqnarray}
S_{\mathrm{BCS}} = N\int d\tau d^{3} x\left( \frac{|\phi|^2}{2\tilde{g}} +  \mathrm{Tr} \ln \mathcal{K}_{\mathrm{den(BCS)}}\right) \ .
\end{eqnarray}
At large $N$, the partition function will be dominated by a mean field solution, $\phi_0$, determined through minimization of 
\begin{eqnarray}
S_{\mathrm{BCS}}  &=&  NVT \left(\int d\omega\int \frac{d^3k}{(2\pi)^3} \ln \left(\omega^2 + k^4/(2M)^2 + \phi_0^2\right)\right. \cr 
&&+\left.\frac{\phi_0^2}{2\tilde{g}}\right) \cr
&=& NVT \left(-2/5\sqrt{2}\Gamma(3/4)M^{3/2}(\phi_0/\pi)^{5/2} + \phi_0^2/2\tilde{g}\right) \ , \cr
\end{eqnarray}
where the final expression follows from that in Refs.~\cite{Lawrence_2023,Romatschke:2023ztk}, using dimensional regularization and identities from Ref.~\cite{Nishida_2007}. Minimization of this action admits two solutions: $\phi_0 = 0$, and the non-trivial,
\begin{eqnarray}
\label{eq:vac}    
\phi_0 = 32\pi [\Gamma(5/4)]^4/(\tilde{g}^2 M^3) \ . 
\end{eqnarray}
Note that near the non-interacting limit, this second solution is pushed to positive infinity. We posit that the trivial solution corresponds to physics near the non-interacting fixed point, while near the unitary fixed point the non-trivial solution is preferred. Indeed, for any $\tilde{g}>0$, the non-trivial solution gives the least action. 

\subsection{Distribution of the 2-fermion correlation function}

On a given background configuration, $\phi$, the two-fermion correlation function is given by,
\begin{eqnarray}
C_2(\phi,\tau) \equiv M\int \frac{d\omega}{2\pi}\frac{e^{-i\omega \tau}}{\omega^2 + \phi^2} = \frac{M e^{-\phi\tau}}{\phi} \ .
\end{eqnarray} 
Turning now to the characteristic function, Eq.~\ref{eq:charfun}, we find,
\begin{eqnarray}
\Phi_{\ln C_2}(s) &=& \mathcal{N}\int \mathcal{D}\phi e^{-S_\phi + i s \ln C_2[\phi,\tau]} \cr
&\equiv& \mathcal{N}\int \mathcal{D}\Phi e^{-S_{\Phi,s}}\ ,
\end{eqnarray}
where in the last line we expand around the mean field solution of the unmodified action, $\phi \equiv \Phi -\phi_0$, of Eq.~\ref{eq:vac}, and identify 
\begin{widetext}
\begin{eqnarray}
\label{eq:effact}
S_{s,\Phi} = NVT \left(-2/5\sqrt{2}[\Gamma(3/4)]^2 M^{3/2}\pi^{-5/2}(\Phi-\phi_0)^{5/2} + (\Phi-\phi_0)^2/(2\tilde{g})\right) + i s (\Phi-\phi_0)T\ ,
\end{eqnarray}
\end{widetext}
where we have taken $\tau=T$, and $\Phi T \gg \ln M/\Phi$. This action describes a new field theory, having propagators proportional to $1/N$, arbitrary $n$-point vertices proportional to $N$, and an external vertex proportional to $s$. Note that the $n$th cumulant is proportional to $s^n$, and therefore contributing diagrams will contain $n$ of these external vertices.

A generic diagram contributing to the $n$th cumulant will therefore scale as $N^{v-P}$, where $v$ is the number of internal vertices and $P$ the number of $\phi$ field propagators. Identifying $V=v+n$ as the total number of vertices, and using the topological invariant $L+V-P=1$, where $L$ is the number of loops, we find that the scaling can be written, $N^{-n+1-L}$. Thus, at leading order in our $N$-counting, only tree-level diagrams contribute, allowing us to use a classical, mean-field solution for $\Phi$ at this order.

 Minimizing the effective action, Eq.~\ref{eq:effact}, $\left. \frac{\partial S_{s,\Phi}}{\partial \Phi}\right|_{\Phi_0} = 0 $ leads to the solution,
\begin{widetext}
\begin{eqnarray}
\Phi_0/\phi_0 &=& \frac{2}{3}\frac{1}{3(NV)^{2/3}\pi^{10/3}}\left[\left((NV)^2\pi^{10}+18i(M \tilde{g})^3 NV\pi^5 s\Gamma[3/4]^4-54(M\tilde{g})^6s^2\Gamma[3/4]^8\right.\right.\cr
&+&\left.\left.6\sqrt{3}\Gamma[3/4]^6\sqrt{(m\tilde{g})^9s^3\left(-2iNV\pi^5+27(M\tilde{g})^3 s\Gamma[3/4]^4\right)}\right)^{1/3}
\right.\cr
&+&\left(\pi^{10/3}\left(NV\pi^5+12i(M\tilde{g})^3 s\Gamma[3/4]^4\right)\right)/\left(\tilde{g}^2\left(NV\left((NV)^2\pi^10+18i(M\tilde{g})^3NV\pi^5 s\Gamma[3/4]^4-54(M\tilde{g})^6 s^2\Gamma[3/4]^8\right.\right.\right.\cr
&+&\left.\left.\left.\left.6\sqrt{3}\Gamma[3/4]^6\sqrt{(M\tilde{g})^9s^2\left(-2iNV\pi^5+27(M\tilde{g})^3s\Gamma[3/4]^4\right)}\right)\right)^{1/3}\right)\right]\cr
&=& 
\frac{1}{4} \sum_{n=0}^{\infty}\frac{a_n}{n!}\left(\frac{-is(M/\tilde{g})^3}{4\pi \Gamma[1/4]^4 NV}\right)^{n+1} \ ,
\end{eqnarray}
\end{widetext}
where the $a_n$ are a set of integer coefficients.

The action is then, 
\begin{eqnarray}
S_{s,\Phi_0} &=& \frac{\tilde{g}T}{4NV}\sum_{n=0}^{\infty}\frac{b_n}{n!}\left(\frac{-8 (M \tilde{g})^3}{4\pi \Gamma[1/4]^4 NV}\right)^{n+1} (is)^n\ ,
\end{eqnarray}
with integer coefficients, $b_n$.
We can determine the cumulants by identifying,
\begin{eqnarray}
\left. n! \frac{\partial^n S_{s,\Phi_0}}{\partial s^n}\right|_{s\to 0} = \kappa_n \ ,
\end{eqnarray}
which gives the following expression for the cumulants:
\begin{eqnarray}
\kappa_n \propto \frac{g T}{V N}\xi^{n-2} \ ,
\end{eqnarray}
where
\begin{eqnarray}
\xi &\equiv& \frac{(M \tilde{g})^3}{\pi \Gamma(1/4)^4N V}\cr
&=& \frac{64\pi^2 (M\hat{g})^3}{ \Gamma(1/4)^4N E_3^{3/2}V} \propto \frac{ (M\hat{g})^3}{ N \Lambda^3V}\ ,
\end{eqnarray}
Thus, to leading order in a $1/N$ expansion, $\kappa_{n>2} = 0$, and this distribution is log-normal. Corrections to log-normal may be determined by computing one loop and higher contributions. 

\section{Conclusions and continuing studies}

Two exploratory calculations have been presented with the intent of refining and potentially expanding the predictions of Ref.~\cite{Nicholson:2012zp}. First, systematic effects due to finite lattice sizes are estimated numerically, and removed through an extrapolation to the continuum limit. We find that corrections to the initial predictions of Ref.~\cite{Nicholson:2012zp} are likely small, but these studies must be expanded and refined. We are currently performing computations including a non-zero explicit three-body interaction, to add further confidence that we are within the universal regime. We are also actively investigating the use of machine learning techniques to translate numerical deviations from a log-normal distribution into corrections of the spectrum.

We have also laid out a path toward an analytic tool for computing these corrections, as a perturbative expansion in the limit where the number of identical species, $N$, becomes large. It is clear that, using an auxiliary field in the density channel of the matter fields, the spectrum is independent of the number of species in the partition function, however, there is no obvious way to perform a perturbative calculation of the probability distribution in the large $N$ limit. In the BCS channel, on the other hand, we are able to leverage large $N$ to our advantage, and find that the leading order form of the probability distribution is log-normal. However, it remains to be shown that the spectrum in this case is unaffected by the superfluous copies of the matter fields. If this is not the case, investigations into corrections due to finite $N$, including one loop and higher contributions to the distribution, may be warranted. 

\section*{Acknowledgments}
The authors would like to thank ECT$^{*}$, and the organizers of the workshop ``Universality in strongly-interacting systems: from QCD to atoms" for their hospitality and engaging conversations on this topic. AN acknowledges P. Romatschke for fruitful discussions on the large $N$ expansion.
AN, JM, and CK were partially supported by the NSF Faculty Early Career Development Program (CAREER) under award PHY-2047185.
JM was supported in part by the U.S. National Science Foundation (NSF) Graduate Research Fellowship Program under Grant
No. DGE-2040435. Any opinions, findings, and conclusions or recommendations expressed in this material are those of the author(s) and do not necessarily reflect the views of the NSF. JM was also supported in part by the U.S. Department of Energy (DOE), Office of Science, Office of Nuclear Physics, under grant contract numbers DE-AC02-05CH11231, the DOE Topical Collaboration “Nuclear Theory for New Physics”, award No. DE-SC0023663, and the U.S. DOE, Office of Science, Office of Workforce Development for Teachers and Scientists, Office of Science Graduate Student Research (SCGSR) program. The SCGSR program is administered by the Oak Ridge Institute for Science and Education (ORISE) for the DOE. ORISE is managed by ORAU under contract number DESC0014664. TR was supported by the NSF through cooperate agreement 2020275 and by the DOE Topical Collaboration ``Nuclear Theory for New Physics", award No. DE-SC0023663.

\clearpage
\bibliography{LN}

\end{document}

%% file: preamble.tex
\usepackage{graphicx}  % needed for figures
\usepackage{dcolumn}   % needed for some tables
\usepackage{bm}        % for math
\usepackage{amssymb}   % for math
\usepackage{standalone}
\usepackage{enumitem}
\usepackage[pdftex]{color}
\usepackage[utf8]{inputenc}
\usepackage{xcolor}
\usepackage{slashed}
\usepackage{booktabs}
\usepackage{multirow}
\usepackage{amsmath}
\usepackage{bbm}
\usepackage{stackrel}
\usepackage{rotating}
\usepackage{CJKutf8}
\usepackage{pifont}
\usepackage{mathtools}
\usepackage[caption=false]{subfig}
\usepackage{listings}
\usepackage{hyperref}
\hypersetup{
    colorlinks=true,       % false: boxed links; true: colored links
    linkcolor=blue,          % color of internal links
    citecolor=blue,        % color of links to bibliography
    filecolor=blue,      % color of file links
    urlcolor=blue           % color of external links
}
\usepackage{simplewick}
\usepackage{float} % Forces placement of figures
\usepackage{tikz} % adds /foreach (looping) command
\usepackage{xspace}
\usepackage[normalem]{ulem}

\usepackage{comment}

\allowdisplaybreaks

\AtBeginDocument{%
    \newwrite\bibnotes
    \def\bibnotesext{Notes.bib}
    \immediate\openout\bibnotes=\jobname\bibnotesext
    \immediate\write\bibnotes{@CONTROL{REVTEX41Control}}
    \immediate\write\bibnotes{@CONTROL{%
    apsrev41Control,author="08",editor="1",pages="1",title="0",year="1"}}
     \if@filesw
     \immediate\write\@auxout{\string\citation{apsrev41Control}}%
    \fi
}%

%% file: def.tex
\usepackage{dsfont}

\def\eqref#1{{(\ref{#1})}}

\definecolor{kngrey}{HTML}{A6AAA9}
\definecolor{knred}{HTML}{EC5D57}
\definecolor{knorange}{HTML}{F39019}
\definecolor{knyellow}{HTML}{F5D328}
\definecolor{kngreen}{HTML}{70BF41}
\definecolor{knblue}{HTML}{51A7F9}
\definecolor{knpurple}{HTML}{B36AE2}

\newcommand{\be}{\begin{equation}}
\newcommand{\ee}{\end{equation}}
\newcommand{\bee}{\begin{equation*}}
\newcommand{\eee}{\end{equation*}}
\newcommand{\bea}{\begin{eqnarray}}
\newcommand{\eea}{\end{eqnarray}}
\newcommand{\beaa}{\begin{eqnarray*}}
\newcommand{\eeaa}{\end{eqnarray*}}

%% file: affiliations.tex
\newcommand{\unc}{
	Department of Physics and Astronomy,
	University of North Carolina,
	Chapel Hill, NC 27516-3255, USA
	}
\newcommand{\lblnsd}{
    Nuclear Science Division,
    Lawrence Berkeley National Laboratory,
	Berkeley, CA 94720, USA
}
\newcommand{\yale}{
	Department of Physics,
	Yale University,
	New Haven, CT 06520 USA
	}
\newcommand{\umd}{
	Department of Physics, 
    University of Maryland, 
    College Park, MD, 20742 USA
	}
\newcommand{\ucb}{
	Department of Physics, 
    University of California, 
    Berkeley, CA 94720, USA
	}